\documentclass{article}  
\usepackage[hang,bf]{caption}  
  
\begin{document}  
  
\title{Role of Coulomb correlation on magnetic and
transport properties of doped manganites: 
La$_{0.5}$Sr$_{0.5}$MnO$_3$ and LaSr$_2$Mn$_2$O$_7$}  
  
\author
{Julia E. Medvedeva\(^ {1,}\)\(^{2,}\)\thanks{Corresponding author.
Fax: +1-847-491-5082. {\it E-mail address:} jem@pluto.phys.nwu.edu} ,  
Vladimir I. Anisimov\(^{2}\), \\  
Oleg N. Mryasov\(^{3}\), Arthur J. Freeman\(^{1}\) \\  
\(^1\) Department of Physics and Astronomy, Northwestern University, \\ 
Evanston, IL 60208-3112 USA \\  
\(^2\) Institute of Metal Physics, Yekaterinburg 620219, Russia \\
\(^3\) Seagate Research, 
Pittsburgh, PA 15203 USA
}  
  
\maketitle  
  
\begin{abstract}  
Results of LSDA and LSDA+U calculations of the electronic structure 
and magnetic configurations of the 50 \% hole-doped 
pseudocubic perovskite La$_{0.5}$Sr$_{0.5}$MnO$_3$ and
double layered LaSr$_2$Mn$_2$O$_7$ are presented. 
We demonstrate that the on-site Coulomb correlation (U) 
of Mn $d$ electrons has a very different influence on the
{\it (i)} band formations, {\it (ii)} magnetic ground states,
{\it (iii)} interlayer exchange interactions, and {\it (iv)} 
anisotropy of the electrical transport in these two manganites.
A possible reason why the LSDA failures in predicting
observed magnetic and transport properties of the double layered
compound - in contrast to the doped perovskite manganite - 
is considered on the basis of a p-d hybridization analysis.
\end{abstract}  

pacs: 71.20.-b, 72.25.-b, 75.30.Kz, 75.50.Cc, 75.50.Ee



\section{Introduction}

The half-doped manganites, La$_{0.5}$Sr$_{0.5}$MnO$_3$ and 
LaSr$_2$Mn$_2$O$_7$, belong to the Ruddlesden-Popper family
\cite{Ruddlesden57},
(R,A)$_{n+1}$Mn$_n$O$_{3n+1}$, which has been attracting intensive 
interest due to itinerant doped carriers that give rise to a variety 
of interesting physical phenomena, notably colossal magnetoresistance.
One of the most striking properties of the double layered manganites,
La$_{2-2x}$Sr$_{1+2x}$Mn$_2$O$_7$, 
is the extremely large magnetoresistance which is much larger
than that for prototypical perovskite systems:
it was found in 40 \% doped La$_{2-2x}$Sr$_{1+2x}$Mn$_2$O$_7$ that
the resistivity ratio reaches $\sim$ 20,000 \% at 7~T above T$_c$ 
\cite{Moritomo96}, whereas for La$_{0.6}$Sr$_{0.4}$MnO$_3$
 $\rho(0)/\rho(H) \sim$ 200 \% at 15~T \cite{Urushibara95}.
The main difference between 3D perovskite and 2D layered compounds 
arises from the change of dimensionality (n) 
of the crystal structure connected with the number of MnO$_6$ layers 
along the $c$ direction (n=$\infty$ for pseudocubic
perovskite systems and n=2 for double layered manganites),
which in turn has a strong influence on their magnetic and 
transport properties
\cite{Moritomo96,Kimura96,Perring97,Kimura98,mag:corr:98}.
Measurements and comparison of characteristic features
(e.g., the strong anisotropy of electrical (magneto-)
transport and magnetostriction, and the
2D character of magnetism caused by reduction 
of the exchange coupling between the Mn ions along the $c$ direction 
in the double-layered manganites)
suggested that both the anisotropic transfer interaction and the
two dimensional spin correlation have a strong influence on
the CMR properties in layered manganites \cite{Moritomo96}.

The pseudocubic perovskites, La$_{1-x}$Sr$_x$MnO$_3$, 
have an insulating paramagnetic phase at high
temperature and a metallic ferromagnetic phase at low temperature 
over a wide range of hole doping \cite{Urushibara95,Sundaresan98}. 
The strong coupling between the magnetic
ordering and the electrical conductivity demonstrates the strong relationship
between the electrical resistivity and the spin alignment
which has been explaind by the double-exchange mechanism arising from
the mixed valence of Mn$^{3+}$/Mn$^{4+}$.
In comparison to these pseudocubic perovskites, 
the double layered structures (La$_{2-2x}$Sr$_{1+2x}$Mn$_2$O$_7$) 
with their two-dimensional networks of MnO$_6$ octahedra  
have a reduced exchange coupling between the Mn ions 
along the {\it c} direction. Indeed, in the double-layered case  
consisting of two perovskite blocks separated by an intervening 
insulating layer of (La,Sr)O ions along the {\it c} axis, 
the balance between antiferromagnetism and ferromagnetism is very 
sensitive to $e_g$ band filling \cite{Battle97}.  
Bilayer La$_{2-2x}$Sr$_{1+2x}$Mn$_2$O$_7$ demonstrates ferromagnetism 
in a doping range $x<0.39$, a canted antiferromagnetic structure 
for a hole concentration $0.39<x<0.48$, and layered antiferromagnetic
states for $x>0.48$ \cite{Kubota99a,Moritomo98}.

An important issue in the theory of CMR oxides is the role
of Coulomb correlation since an accurate treatment of correlation may affect
significantly the balance between ferromagnetic and antiferromagnetic
interactions and hence the magnetic ground state.
According to our previous investigations \cite{Medvedeva01},
where we demonstrated that the on-site Coulomb correlation of Mn d electrons
significantly modifies the electronic structure, magnetic ordering,
interlayered exchange interactions and promotes strong
electronic transport anisotropy in 50 \% doped LaSr$_2$Mn$_2$O$_7$,
the experimentally observed magnetic ordering is reproduced within
LSDA+U calculations only for U$\geq$7 eV.
This is in contrast with the CMR perovskite 3D manganites
(LaMnO$_3$, (La,Sr)MnO$_3$ systems) where LSDA theory correctly predicts
the magnetic ground state \cite{Pickett96,Solovyev00}.

Considering two 50 \% doped manganites, the double layered 
LaSr$_2$Mn$_2$O$_7$ and pseudocubic perovskite La$_{0.5}$Sr$_{0.5}$MnO$_3$, 
we outline here the main similarities
and differences in their characteristic properties:
{\it (i)} Both manganites formally consist of Mn$^{3.5+}$ ions
surrounded by oxygen octahedra, and so in a high spin configuration
the majority $t_{2g}$ states are filled and the electrically active orbitals 
are $d_{x^2-y^2}$ and $d_{3z^2-r^2}$.
{\it (ii)} The dynamical Jahn-Teller (JT) effect is absent 
for both systems, $\Delta_{JT}=D(Mn-O_{apical})/D(Mn-O_{planar})$, where 
$D(Mn-O)$ is manganese oxygen distance, equals to 1.004 and 1.001 
for La$_{0.5}$Sr$_{0.5}$MnO$_3$ and LaSr$_2$Mn$_2$O$_7$, respectively --
although it should be noted here that due to an intervening 
layer of (La.Sr)O ions along the $c$ axis in the latter manganite, 
its MnO$_6$ octahedra are slightly distorted
in terms of different distances between Mn and apical oxygens.
Hence, in contrast to the 3D perovskite manganites which have
two types of oxygen atoms (one apical and one planar), there are three oxygen
types in the double layered manganites -- two apical and one planar.
{\it (iii)} The resistivity of La$_{0.5}$Sr$_{0.5}$MnO$_3$ is 
of the order of 10$^{-3} \Omega cm$ with a flat temperature dependence
\cite{Sundaresan98}, whereas for LaSr$_2$Mn$_2$O$_7$ 
the $ab$-plane resistivity is 10$^0$--10$^{-1} \Omega cm$ \cite{Kubota99}
and the resistivity perpendicular to the layer is $\sim$10$^2$ times larger;
in addition, according to previous theoretical works 
\cite{Solovyev00,Medvedeva01,Huang00} both are half-metals.
{\it (iv)} They have different magnetic orderings:
La$_{0.5}$Sr$_{0.5}$MnO$_3$ is a ferromagnet (FM), while
LaSr$_2$Mn$_2$O$_7$ is an A-type antiferromagnet (AFM)
(magnetic moments lie in the $ab$ plane and couple ferromagnetically
within the single MnO$_2$ layer, but show AFM order between layers). 
{\it (v)} As mentioned above, the different influence of
the on-site Coulomb correlation on the magnetic ordering in 
these manganites has been established in the previous calculations.

In this paper, on the basis of comprehensive analyses of the crystal
structures, 
electronic structures and magnetic configurations of the double layered 
LaSr$_2$Mn$_2$O$_7$ and pseudocubic perovskite La$_{0.5}$Sr$_{0.5}$MnO$_3$ 
compounds, we establish the mechanisms governing the band formation
in these manganites and the different role of the on-site Coulomb correlation
in its relation with the electronic, magnetic and transport properties.
For both compounds, we consider the total energy differences between FM 
and AFM spin configurations for the Coulomb correlation parameter, U,
equal to 0 and 7 eV; then, for the ground states of these manganites,
we compare the exchange interaction parameters and the most important
band characteristics (such as the population of Mn-d states 
near the Fermi level, Mn-d sub-band widths and the p-d hybridization) 
and give a possible reason why the LSDA scheme (U=0) fails to
predict the magnetic ground state as well as the 
electrical transport properties for the double layered manganite,
in contrast to its success for the case of the pseudocubic perovskite
manganite.

\section{Methodology}
  
The LSDA and LSDA+U calculations were performed in the frame-work of the   
linear-muffin-tin-orbital method in the atomic sphere approximation  
(LMTO-ASA) \cite{Andersen75}. We used the von Barth-Hedin-Janak form 
\cite{Barth75} for the exchange-correlation potential.  
The crystal parameters were taken from Ref. \cite{Woodward98,Argyriou00}.  
From the constrained LSDA supercell calculations 
\cite{Gunnarsson89,Anisimov91},  
we obtained values of the Coulomb and exchange parameters to be
U=7.1 eV and J=0.79 eV for La$_{0.5}$Sr$_{0.5}$MnO$_3$ and 
U=7.2 eV and J=0.78 eV for LaSr$_2$Mn$_2$O$_7$. 
These values are typical for the transition-metal 
oxides \cite{ldau,Satpathy96,Solovyev96}.    

\section{Results and Discussion}
\subsection{Electronic structure, transport anisotropy and magnetism}
  
As a first step, we performed the band structure calculations   
of the FM (all atoms in every layer and between layers are ordered 
ferromagnetically) and the A-type AFM (ferromagnetic layers stacked 
antiferromagnetically) phases of both La$_{0.5}$Sr$_{0.5}$MnO$_3$ and 
LaSr$_2$Mn$_2$O$_7$ by the standard LSDA method.
FM spin alignment is found to be more preferable from total
energy calculations for both manganites (Table 1).
The projected densities of states (DOS) for these FM
cases look very similar (Fig. 1): For the majority-spin channel, 
Mn 3d states of both compounds
form the bands between 2.2 eV below E$_F$ and 2.5 eV 
above E$_F$. As seen from the figure, the $e_g$ ($x^2-y^2$ 
and $3z^2-r^2$) bands are partially filled, cross E$_F$, and are rather 
broad compared to the $t_{2g}$ ($xy$ and degenerate $xz$, $yz$) bands
which are about 1.2 eV wide and lie 1 eV below E$_F$. 
The exchange interaction splits 
the Mn 3d states such that the $t_{2g}$ and $e_g$ minority-spin bands 
are located 2.6 eV higher in energy than the  majority-spin states.  
We obtained a band gap in the minority-spin $d$ bands  
of around 2.5 eV for both La$_{0.5}$Sr$_{0.5}$MnO$_3$ and 
LaSr$_2$Mn$_2$O$_7$ -- in keeping with the fact that
doped perovskite manganites are found  
to be half-metallic \cite{Wei97,Park98,Solovyev00}.
The analogs in the main band characteristics near 
E$_F$ in the calculated electronic spectra of the double layered  
and doped perovskite manganites allow us to suggest
that the insulating character of minority-spin electrons
for the double-layered manganite
may also render the conduction band a half-metallic one 
(i.e., 100 \% spin polarization for the conduction electrons).
Note that $E_F$ in our
spin-polarized band structure calculations of the double-layered
manganite was found to lie at the bottom of the $t_{2g}$ minority 
spin conduction band 
(which is in agreement with the full-potential 
LMTO calculations for LaSr$_2$Mn$_2$O$_7$ \cite{Huang00}), and
so the LSDA electronic band structure just misses to being half-metallic.

The crucial role of Coulomb correlation
in the stability of the experimentally observed magnetic ground state
of LaSr$_2$Mn$_2$O$_7$ was established in Ref. \cite{Medvedeva01}.
We investigated the electronic structure similarities 
of LaSr$_2$Mn$_2$O$_7$ and La$_{0.5}$Sr$_{0.5}$MnO$_3$  
for both FM and A-type AFM spin alignments in the frame-work of 
the LSDA+U method with the calculated U value of 7 eV.
According to the total energy differences in Table 1,
the ground state of the perovskite manganite remains ferromagnetic,
while the observed A-type AFM ordering in double layered manganite
is found to be energetically more favorable with U=7 eV.
(A possible reason for the LSDA failure for the case of 
the double layered manganite in contrast to the perovskite
will be discussed later). A comparison of the projected DOS's for 
the LSDA+U ground states of La$_{0.5}$Sr$_{0.5}$MnO$_3$ (solid line) 
and LaSr$_2$Mn$_2$O$_7$ (dashed line) is shown in Fig. 2.
Only majority spin $e_g$ states contribute to the DOS at E$_F$.
We find the magnetic moment of the primitive unit cell to be equal 
to 7.00 $\mu_B$.  
Thus, LSDA+U calculations result in a truly half-metallic state 
for the double layered manganite, and so
electron conduction is possible only within the majority spin sublattice.

One of the main differences between the doped pseudocubic perovskite 
and double layered manganites which is clearly seen from Fig. 2,
is the different behavior of the electrically active states when 
on-site Coulomb correlation is taken into account. 
The DOSs for the LaSr$_2$Mn$_2$O$_7$ $e_g$ states demonstrate (Fig. 2)
the significant difference between $3z^2-r^2$ and $x^2-y^2$ orbitals in
the electron population near $E_F$, while for 
La$_{0.5}$Sr$_{0.5}$MnO$_3$ the band widths and the position
of the band centers of these two $e_g$ orbitals as well as their 
contributions to the DOS at $E_F$ are very close (Fig. 2).

In Table 1 we presented the ratio between the $x^2-y^2$ 
and $3z^2-r^2$ contributions to the DOS at E$_F$ 
(denoted as N$_{x^2-y^2}$/N$_{3z^2-r^2}$) calculated in both 
the LSDA and LSDA+U approaches. 
The ratio demonstrates the presence of anisotropy in the 
population of the two $e_g$ states at E$_F$; and
since the situation does not change qualitativly 
in the 0.3 eV energy window just above E$_F$ (Fig. 2), 
we draw the following conclusions in the picture of 
charge transport: For La$_{0.5}$Sr$_{0.5}$MnO$_3$, 
we have 3D conduction; the use of U=7 eV does not result 
in a large splitting of the $e_g$ states
and the number of electrons that contribute to electrical
transport remain almost insensitive to the influence of U.
This is contrary to that in LaSr$_2$Mn$_2$O$_7$ where for U=7 eV
the electron conduction (hopping) along the $c$ axis 
(N$_{3z^2-r^2}$) becomes very small and thus, 
on-site correlation treated within LSDA+U promotes 2D type 
(in-plane) electronic behavior in the double layered system
--- as observed experimentally.
Thus, the comparison of the LSDA and LSDA+U band structures 
for both manganites leads us to the following conclusion:
the exchange splitting with almost unoccupied minority 
spin Mn-3d states and the ligand field splitting of the $t_{2g}$ and 
$e_g$ states are the main factors governing band formation in both 
La$_{0.5}$Sr$_{0.5}$MnO$_3$ and LaSr$_2$Mn$_2$O$_7$. 
However, futher splitting of the $e_g$ states  
with increase of U is an essential feature 
of the layered manganites, but is absent in La$_{0.5}$Sr$_{0.5}$MnO$_3$.

The total energy differences for the two manganites 
for U=0 and 7 eV shown in Table 1 (where the total energy 
of the FM spin ordering for each U value is taken to be zero)
demonstrate that for the double layered compound the on-site Mn-d 
electron Coulomb correlations modify the magnetic ordering 
from FM to A-type AFM -- as experimentally observed, while
the FM ground state of the pseudocubic perovskite remains 
energetically more preferred as U is increased. 
To determine the character of the exchange coupling in these manganites,
we calculated the effective exchange 
interaction parameters (for computational details see \cite{lichtan}).
In Table 1, the values of the d-d effective exchange parameters, 
$J_{dd}$, between nearest Mn neighbours which belong to the different 
layers of one bilayer are shown for both U values.
As seen from the Table, they follow the magnetic ordering behavior 
(obtained by the total energy differences between FM and AFM configurations)
for both compounds. (For more details of the change of the 
exchange interaction parameters and the FM-AFM total energy differences
in LaSr$_2$Mn$_2$O$_7$ with the increase of U, see Ref. 
\cite{Medvedeva01}.)

We interpreted the calculated exchange coupling within 
the superexchange (SEX) and double exchange (DEX) models by calculating
the $e_g$-sub-band widths of the manganite ground states. 
As can be seen from Table 1, where we present the ratio
of the $x^2-y^2$ and $3z^2-r^2$ band widths
(denoted as W$_{x^2-y^2}$/W$_{3z^2-r^2}$), 
for La$_{0.5}$Sr$_{0.5}$MnO$_3$ the $e_g$-band widths 
stay exactly equal with increase of U, and provide positive DEX interactions.
We note here, that the increase of the ferromagnetic 
contributions to the exchange interaction energy (c.f., Table 1) 
may be due to  
the increase of the $e_g$-band widths from 3.4 eV for U=0 
to 3.8 eV for U=7 eV and the stronger O-p hybridization
in the Mn-d states -- as will be illustrated below. 
In LaSr$_2$Mn$_2$O$_7$, a sharp increase of the W$_{x^2-y^2}$/W$_{3z^2-r^2}$ 
ratio coming from the narrowing of the $3z^2-r^2$ band width 
(Table 1 and Fig. 2)
results in the strong splitting of these two $e_g$-sub-bands and so
promotes the negative (AFM) contribution to the exchange interaction.
Thus, Coulomb correlations have a very different influence 
on pseudocubic and double layered manganites: for the first
we obtain an increase of the DEX contributions, while for
LaSr$_2$Mn$_2$O$_7$, U supresseses the DEX contributions 
to the interlayer exchange interaction energy and
produces the change in magnetic ordering from FM to AFM. 

\subsection{Significance of p-d hybridization}

The degree of O-p hybridization with Mn-d states plays an important 
role in the band structure and on the exchange mechanism 
\cite{Pickett96,Koizumi01}. 
We compared below the p-d hybridization of La$_{0.5}$Sr$_{0.5}$MnO$_3$ 
and LaSr$_2$Mn$_2$O$_7$ for U=0 and 7 eV 
by calculating the O-p contributions to the DOS
at the energy windows of corresponding Mn-d states. 
And the comparison gives interesting differences between double layered 
and perovskite manganites, supporting the results on magnetic
and transport properties we have already discussed above. 

According to the qualitative picture of the simple DEX model,
an electrically active electron has a finite hopping probability (t)
between ferromagnetically ordered Mn ions, but this hopping vanishes
in the case of antiferromagnetic spin alignment due to strong Hund coupling
(t$<<$J). The character of the occupied $e_g$ orbitals, 
which affects transport and magnetic properties, has been established 
above, but it is also important to investigate the role of
oxygen hybridization in the $e_g$ and $t_{2g}$ orbitals as well 
\cite{Koizumi01}.

At first, we shall consider a significance of oxygen p-hybridization
in the electrically active $e_g$ orbitals. For both manganite compounds,
two $e_g$ ($3z^2-r^2$ and $x^2-y^2$) and $p_x$, $p_y$, $p_z$ oxygen states
contribute to the DOS at $E_F$. In Table 1, we present
the percentage of these contributions for unit cells 
consisting of two Mn atoms for both compounds.
As expected, only $p_z$ orbitals of apical oxygens
($p_x$, $p_y$ of planar oxygens) hybridize with ``out-of-plane''
$3z^2-r^2$ (``in-plane'' $x^2-y^2$) d-orbitals. 
Comparing the lowest spin configurations within LSDA and 
LSDA+U (as obtained by total energy differences),
we draw the following conclusions:
for ferromagnetic La$_{0.5}$Sr$_{0.5}$MnO$_3$, we have
significant contributions to the DOS at E$_F$ from both apical 
and planar oxygens, and the on-site Coulomb correlation resulting 
in a redistribution of the conduction electrons between $e_g$ 
and p-states, do not change the 3D character of the hopping matrix
(Table 1). The influence of U on the population picture at E$_F$ in 
LaSr$_2$Mn$_2$O$_7$ is very different. 
Established above sharp decrease of the N$_{3z^2-r^2}$ for U=7 eV 
occurs together with vanishing apical oxygen contributions 
at E$_F$ (Table 1) resulting in suppression by U the $c$ axis transport. 
This leads to the superexchange
antiferromagnetic coupling along the $c$ axis --
as was obtained by calculating the interlayer exchange interaction
parameters (Table 1). In contrast,
a sharp increase of the $p_x$ ($p_y$) hybridization in the
$x^2-y^2$ orbital (due to the charge redistribution) 
provides strong DEX ferromagnetic interactions in the $ab$-plane.

Investigation of the oxygen contributions to the DOS at E$_F$
allowed us to draw conclusions on the transport and exchange properties
in La$_{0.5}$Sr$_{0.5}$MnO$_3$ and LaSr$_2$Mn$_2$O$_7$. 
Overlapping of Mn-$t_{2g}$ and O-p orbitals in these manganites
is an essential factor governing their band formation.
In order to compare the p- and d-band overlaps for 
the perovskite and double layered compounds, we should,
first of all, compare Mn-O distances for these systems.
The distances between Mn and apical and planar oxygens 
in La$_{0.5}$Sr$_{0.5}$MnO$_3$ are 1.941 and 1.934 \AA, respectively;
for the double layred manganite they are $\sim$0.01 \AA \, smaller 
than for the perovskite compound: 1.917 (1.940) \AA \, for Mn and apical
oxygens and 1.927 \AA \, for Mn and planar oxygen distances. 
Such a negligible difference in Mn-O distances serves to stress 
our results below.

In Fig. 3, we present the calculated angular electron density distribution 
in the energy window of corresponding majority spin $t_{2g}$ states for
both manganites. As seen from the figure, the hybridization 
between Mn-d $t_{2g}$ and O-p states for La$_{0.5}$Sr$_{0.5}$MnO$_3$ 
exists for both U=0 and 7 eV,
becoming stronger as U is increased (Fig. 3(a) and 3(b)). 
The same result was obtained for undoped LaMnO$_3$ \cite{to-be},
in which, however, strong Jahn-Teller distortions of MnO$_6$ octahedra 
lead to more than two times larger oxygen contributions to $t_{2g}$ Mn states 
than that for La$_{0.5}$Sr$_{0.5}$MnO$_3$.
In contrast to these perovskites, and despite the almost equal 
Mn-O distances in the doped manganites,
LSDA band structure calculations for LaSr$_2$Mn$_2$O$_7$ gave 
O-p contributions to the DOS at the energy windows corresponding 
to $t_{2g}$ states that are almost negligible (Fig. 3(c)). 
The p-d hybridization for the double layered
manganite is correctly described only treated in LSDA+U (Fig. 3(d)):
it becomes very strong between Mn and planar oxygens and
one of the apical atoms, O1 (which lies between Mn-layers of one 
MnO$_6$ bilayer unit),
while it is supressed for apical O2 atoms which belong to the 
(La,Sr) insulating layer as well.
Thus, we believe that the absence p-d hybridization 
as an important ingredient in the band formation picture 
results in the failure of the LSDA scheme in predicting 
observed magnetic and transport properties 
of the double layered manganite.

The lowered dimensionality of the magnetic interactions 
in the double layered system, which was established
under the influence of the on-site Coulomb correlation, 
is also seen from a comparison 
of two LSDA+U cases for La$_{0.5}$Sr$_{0.5}$MnO$_3$ 
and LaSr$_2$Mn$_2$O$_7$ (Fig. 3(b) and 3(d)): 
in the 3D-compound, we have equal contributions
to the DOS from all p-states ($p_x$, $p_y$ and $p_z$) 
of planar oxygens, while for the double layered manganite
only $p_x$($p_y$) states of the planar oxygens 
give significant contributions to the DOS at the energy window 
corresponding to Mn-d $t_{2g}$ states.
In addition, the absence of the hybridization between Mn-d states
and apical O2-p states (this oxygen lies at the boundary of 
the perovskite-type bilayer) 
makes the whole hybridization picture obtained for 
the double layered LaSr$_2$Mn$_2$O$_7$ for U=7 eV (Fig. 3(d))
to be in a very good agreement with the structural peculiarity 
of this compound - namely the existance of two perovskite-type layers
separated by a rock-salt-type layer (La,Sr)$_2$O$_2$.

\section{Summary}

In summary, the calculated electronic band structure, exchange interaction
parameters, transport properties and p-d hybridization 
of the double layered LaSr$_2$Mn$_2$O$_7$ are significantly modified
under on-site Coulomb correlation U, resulting in the observed magnetic
ground state and transport anisotropy. 
The comparison with the band characteristics of 
the 3D perovskite La$_{0.5}$Sr$_{0.5}$MnO$_3$ sets these similar doped
manganites apart by means of different role of the Coulomb correlation on
magnetism and transport properties.

\vspace{1cm}
 
Work at Northwestern University supported by the U.S. Department of Energy
(grant No. DE-F602-88ER45372) and at Institute of Metal Physics supported by
RFFI (grant No. 01-02-17063).

\newpage 
\begin{figure}[t]  
\caption{  
Calculated LSDA Mn-d projected densities of states (in states/eV-unit cell)
of FM La$_{0.5}$Sr$_{0.5}$MnO$_3$ (solid line) and 
LaSr$_2$Mn$_2$O$_7$ (dashed line). The Fermi level is located at 0 eV.
The minority spin contributions are shown as negative states.}  
\end{figure}  
  
\begin{figure}[t]  
\caption{  
The Mn-d projected densities of states (in states/eV-unit cell) 
of FM La$_{0.5}$Sr$_{0.5}$MnO$_3$ (solid line) 
and AFM LaSr$_2$Mn$_2$O$_7$ (dashed line) 
ground states from LSDA+U calculations with U=7 eV.   
The Fermi level is located at 0 eV.
The minority spin contributions are shown as negative states.}  
\end{figure}  
  
\begin{figure}  
\caption{
The calculated angular electron density distribution in the energy window of
the corresponding majority spin $t_{2g}$ states for 
(a) La$_{0.5}$Sr$_{0.5}$MnO$_3$, U=0; 
(b) La$_{0.5}$Sr$_{0.5}$MnO$_3$, U=7;
(c) LaSr$_2$Mn$_2$O$_7$, U=0 eV and 
(d) LaSr$_2$Mn$_2$O$_7$, U=7 eV.
Only one MnO$_6$ octahedron is shown (for the double layered manganite 
it belongs to the upper layer of the bilayer).}  
\end{figure}

\begin{table}
\caption{
Total energy differences, $\Delta E_{TOT}$ in meV; 
the interlayer exchange interaction parameters, $J_{dd}$ in meV; 
the ratio of the $3z^2-r^2$ and $x^2-y^2$ orbital contributions 
to the DOS at E$_F$, N$_{x^2-y^2}$/N$_{3z^2-r^2}$; 
the ratio of the $3z^2-r^2$ and $x^2-y^2$ band widths,
W$_{x^2-y^2}$/W$_{3z^2-r^2}$; the percentage Mn $e_g$ and O-$p$ 
contributions to the DOS at E$_F$, N, 
for La$_{0.5}$Sr$_{0.5}$MnO$_3$ and LaSr$_2$Mn$_2$O$_7$.  
}
\begin{center}
\begin{tabular}{lrrrr} \hline 
 & \multicolumn{2}{c}{
La$_{0.5}$Sr$_{0.5}$MnO$_3$}  
 & \multicolumn{2}{c}{
LaSr$_2$Mn$_2$O$_7$}  \\ 

 &  U=0   &  U=7  &  U=0   &  U=7    \\ \hline 
		  
$\Delta E_{TOT}$ & +338.9 & +473.5 & +91.0 & -112.4  \\ 

$J_{dd}$         & +15.28 & +94.50 & +7.53 & -1.84   \\ 

N$_{x^2-y^2}$/N$_{3z^2-r^2}$ &  1.04  &  0.95  &  0.94 &   3.18  \\ 

W$_{x^2-y^2}$/W$_{3z^2-r^2}$ &  1.00  &  1.00  &  1.24 &   2.38  \\ \hline

N(Mn-$d_{3z^2-r^2}$)&  30  &   24  &  34  &  10    \\

N(Mn-$d_{x^2-y^2}$) &  34  &   22  &  32  &  32   \\

N(O1$_{apical}$(O2$_{apical}$)-$p_z$)&  12  &   18  &   2 (8)  &  0 (0)    \\

N(O3$_{planar}$)-$p_x$/$p_y$) &  24  &   36  &  24  &  58   \\ \hline
\end{tabular}
\end{center}
\end{table}

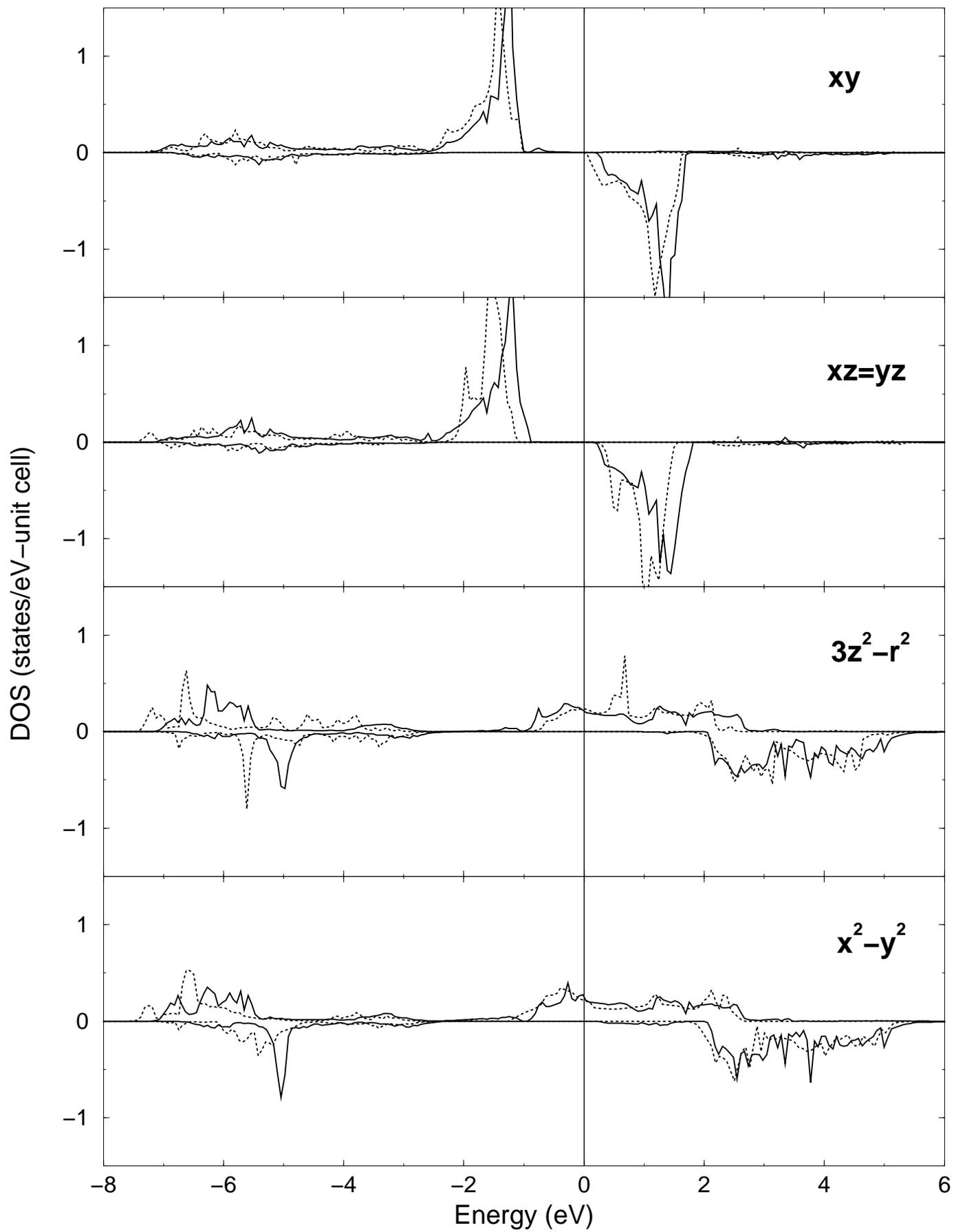
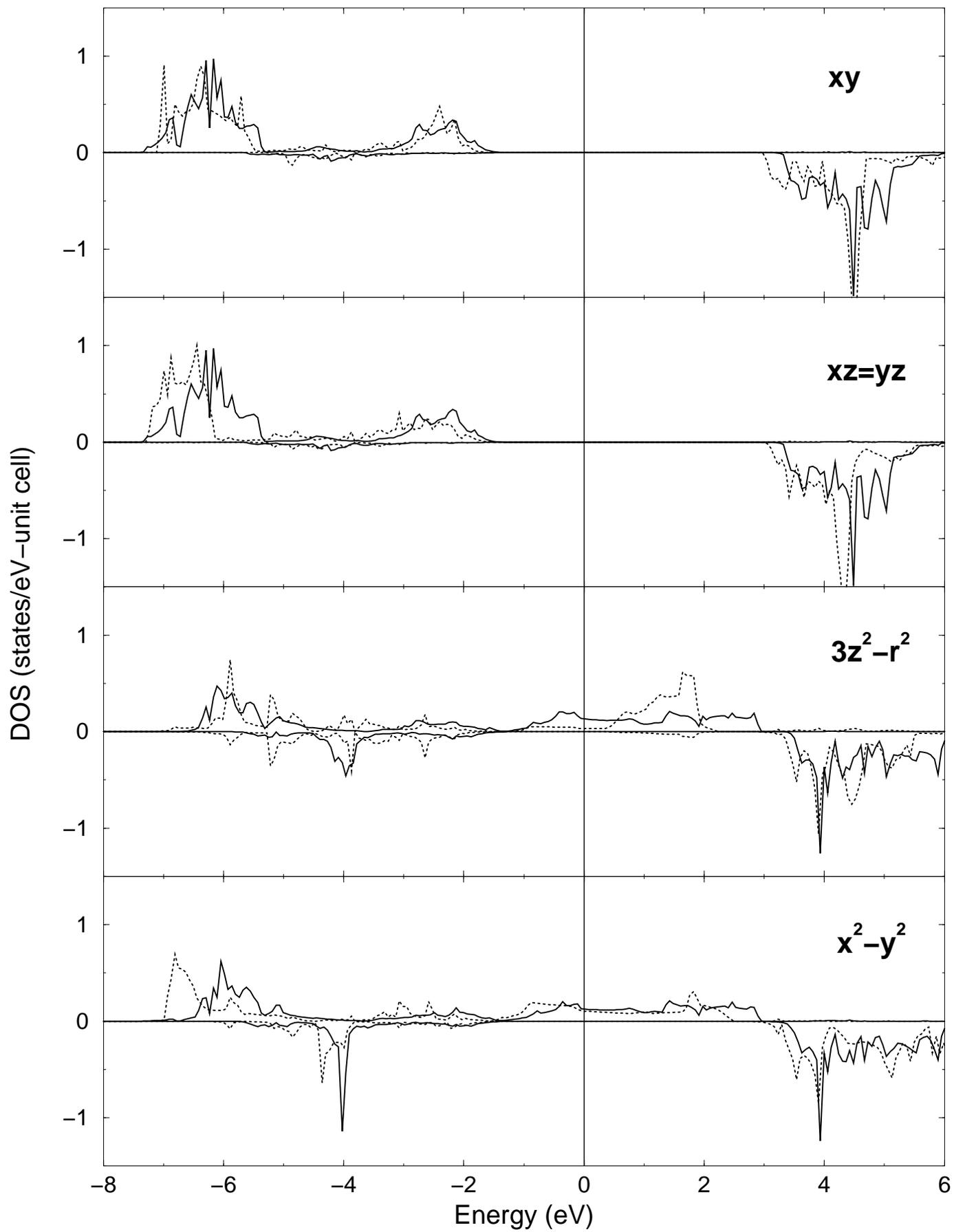
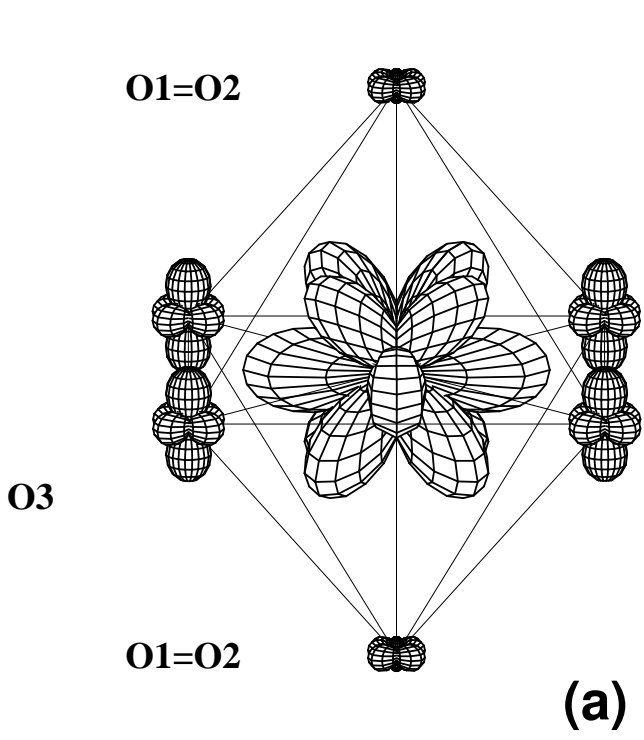
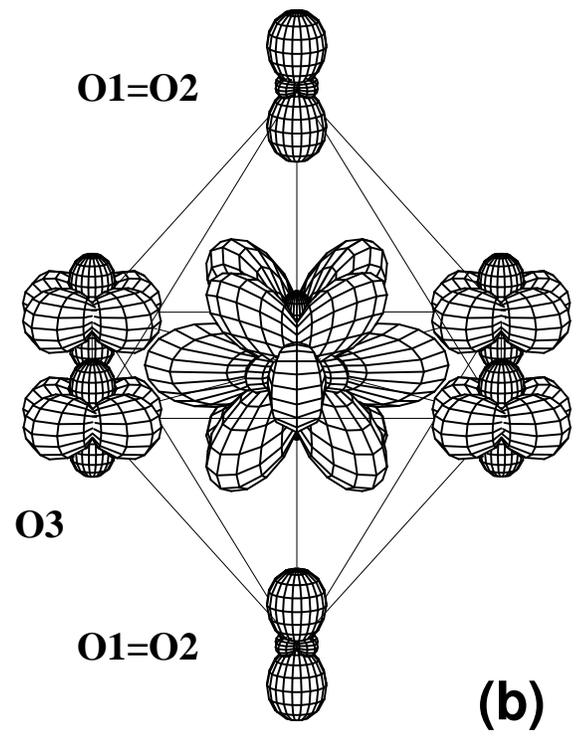
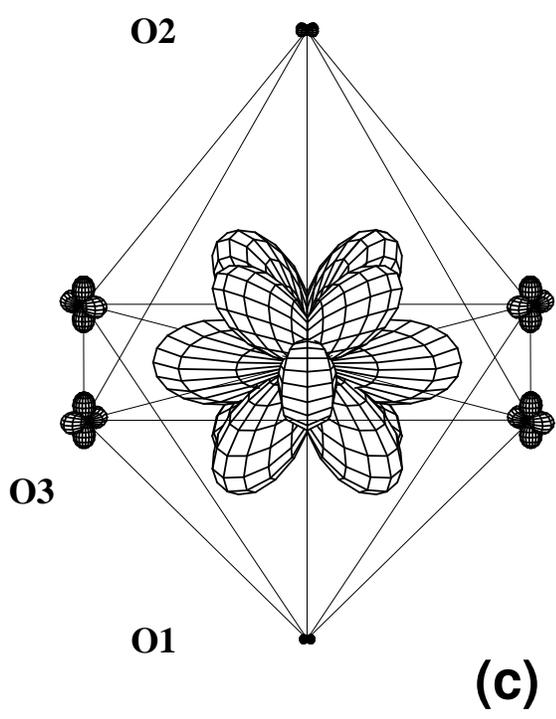
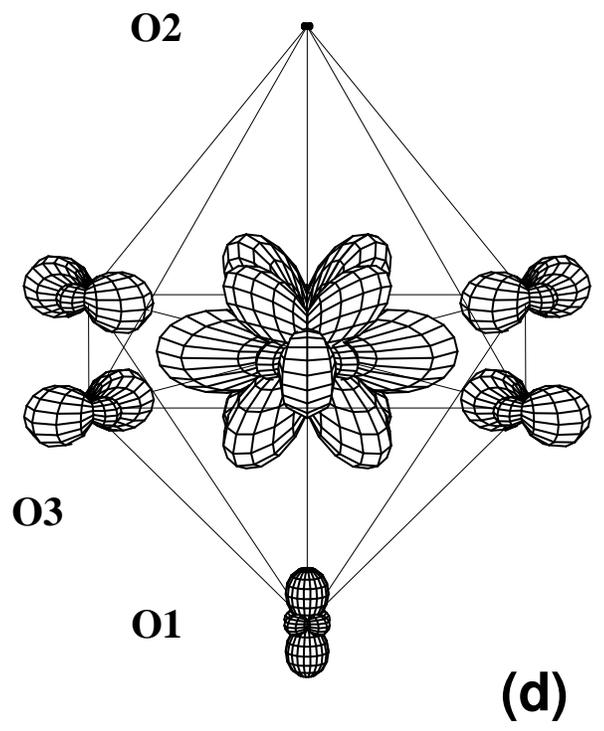
\end{document}